\documentclass[aip,rsi,reprint]{revtex4-2}

\usepackage{newtxtext, newtxmath} 
\usepackage{graphicx} 
\usepackage{hyperref} 
\hypersetup{colorlinks=true,linkcolor=blue,citecolor=blue} 
\usepackage{siunitx} 
\usepackage{mathtools} 
\usepackage{sidecap} 

\newcommand{\nifs}{National Institute for Fusion Science, 322-6 Oroshicho, Toki, Gifu 509-5292, Japan}
\newcommand{\hokudai}{Division of Quantum Science and Engineering, Graduate School of Engineering, Hokkaido University, Kita 13, Nishi 8, Kita-ku, Sapporo, Hokkaido 060-8628, Japan}
\newcommand{\sokendai}{The Graduate University for Advanced Studies (SOKENDAI), 322-6 Oroshicho, Toki, Gifu 509-5292, Japan}
\newcommand{\nagoya}{Graduate School of Science, Nagoya University, Furocho, Nagoya, Aichi 464-8602, Japan}

\begin{document}

\title{Conceptual design of Thomson scattering system with high wavelength resolution in magnetically confined plasmas for electron phase-space measurements}

\author{Kentaro Sakai}
\email[]{sakai.kentaro@nifs.ac.jp}
\affiliation{\nifs}

\author{Kentaro Tomita}
\affiliation{\hokudai}

\author{Takeo Hoshi}
\affiliation{\nifs}
\affiliation{\sokendai}

\author{Akito Nakano}
\affiliation{\nifs}
\affiliation{\sokendai}

\author{Motoshi Goto}
\affiliation{\nifs}
\affiliation{\sokendai}

\author{Kenichi Nagaoka}
\affiliation{\nifs}
\affiliation{\nagoya}

\author{Ryo Yasuhara}
\affiliation{\nifs}
\affiliation{\sokendai}

\date{\today}

\begin{abstract}

We discuss the conceptual design of a spatially-resolved spectroscopy system of Thomson scattering with high wavelength resolution capable of measuring the shape of electron velocity distribution functions in magnetically confined plasmas. We design a spatially-resolved spectrometer with 2560 wavelength channels. The estimated number of scattered photons in a single spectrometer channel is much larger than unity under the experimental setup and plasma parameters of the Compact Helical Device (CHD), indicating sufficient photon statistics for single-shot measurements. Simulations of the scattered spectra show that the signal-to-noise ratio exceeds 5 even under the most unfavorable conditions expected in CHD at full spectral resolution, and further improves with post-processing pixel binning. Bayesian inference applied to the simulated spectra demonstrates that the inferred plasma parameters agree with the input values within the estimated uncertainties. Comparisons between spectra generated from non-Maxwellian electron velocity distribution functions and their Maxwellian fits indicate that deviations from Maxwellian distributions can be identified using the proposed system.

\end{abstract}

\maketitle

\section{Introduction}

In low-collisionality plasmas such as space, solar, and magnetically confined fusion plasmas, non-Maxwellian velocity distribution functions of electrons and ions, including deformations in bulk distribution functions and non-thermal tails, have been observed because the collisional relaxation is ineffective and these non-Maxwellian velocity distribution functions persist over a longer time in comparison with macroscopic/collective time scale \cite{burgess12ssr,amano20prl,jebaraj24apjl,van-lammeren92nf,beausang11rsi,rome97ppcf,mutoh07nf,popov15ppcf,ida22cphys}. 
In magnetically confined fusion plasmas, self-heating by alpha particles, external heating using neutral beams and cyclotron resonances, and Fermi-like stochastic acceleration can lead to non-Maxwellian distribution functions of both electrons and ions in a steady state \cite{pastor12nf,yu24rsi,salewski25nf}. Impulsive energy release events can also generate non-Maxwellian distribution functions as a transient state relaxing toward Maxwellian distribution functions \cite{ida22cphys}. As a specific source of non-Maxwellian electron distribution functions, electron cyclotron resonance heating, for example, can distort the electron distribution function, exhibiting anisotropy with respect to the magnetic field and a superthermal population \cite{rome97ppcf}. The superthermal population of the distribution function can drive radial current and consequently modify the toroidal flow \cite{yamamoto21pop}. Although numerical studies involving kinematic effects, such as gyrokinetic simulations \cite{garbet10nf}, have been widely performed, experimental observations using Thomson scattering \cite{van-lammeren92nf,beausang11rsi} and electron cyclotron emission \cite{yu24rsi} have been limited due to the insufficient resolution of electron distribution functions. 
We are preparing a research project to investigate kinetic effects on macroscopic plasma characteristics with a challenging plasma experiment using the Compact Helical Device (CHD), which is the upgraded device of the Compact Helical System (CHS) \cite{matsuoka88iaea}. The target plasma parameters of CHD are as follows; the maximum magnetic field strength is $B_{max} \sim \SI{1.5}{T}$, the electron temperature is $T_e \sim 100$--\SI{6000}{eV}, and the electron density is $n_e \sim 0.2$--\SI{5e19}{m^{-3}}. We are going to install advanced measurement systems to observe the phase-space structures of electrons and ions using, for example, Thomson scattering, electron cyclotron emission, and charge exchange recombination spectroscopy, together with electric and magnetic field measurements. 
In the Large Helical Device (LHD) experiments, which employ a helical configuration similar to that of CHD, there are several examples showing non-Maxwellian distribution functions. The polarization spectroscopy shows a clear anisotropy of approximately an order of magnitude, and the degree of anisotropy increases as the collision frequency decreases \cite{goto21pfr}. The strong anisotropy may correspond to non-Maxwellian distributions and may also lead to the formation of non-Maxwellian distributions during the isotropization process, for example via instabilities. Thomson scattering measurements performed using a 12-channel polychrometer system show deviations from Maxwellian distribution functions at both high (5--\SI{10}{keV}) and low (200--\SI{1000}{eV}) bulk temperatures \cite{yamada25jfe}. The observed non-Maxwellian distribution functions include bulk deformations and high-energy tails and exhibit both quasi-steady state and transient behavior. These LHD experiments strongly suggest the existence of non-Maxwellian distribution functions under a wide range of plasma conditions, and therefore we need a diagnostic system capable of measuring a variety of non-Maxwellian distribution functions, rather than one that operates only under limited conditions.

Thomson scattering is light scattering by charged particles in the classical limit, where the photon energy is lower than the electron rest mass energy and the light energy is conserved during the scattering process. When light is injected into plasma electrons, the electrons oscillate at the light frequency, and the oscillating electrons emit dipole radiation as scattered light. The frequency of emitted radiation is the same as the incident light frequency with an electron at rest and is Doppler-shifted with a moving electron, i.e., Thomson scattering measurement is Doppler spectroscopy of electrons. With a monochromatic probe beam, the wavelength spectrum of the scattered light is directly related to the shape of the electron velocity distribution function because the observed spectrum is the sum of scattered signals from many electrons within the scattering volume \cite{froula11,hutchinson02}. This explanation is valid with $k\lambda_D \gg 1$, where $k$ is the scattering wavenumber and $\lambda_D$ is the Debye length. Given the plasma parameters in CHD, the lowest $k \lambda_D \sim  250 (T_e/\SI{100}{eV})^{0.5} (n_e/\SI{5e19}{m^{-3}})^{-0.5}$ and the scattering can be incoherent. Assuming a Maxwellian electron velocity distribution function, the electron temperature is estimated from the width of the scattered spectrum. The total scattered intensity is proportional to the electron density. Therefore, Thomson scattering measurement is widely applied to diagnostics of electron temperature and density in a variety of plasmas \cite{narihara95rsi,narihara01rsi,pasqualotto04rsi,van-der-meiden06rsi,lee10rsi,pasch16rsi,sakai22srep,hassaballa04iiie}, as well as non-Maxwellian velocity distribution functions such as fast electron beams in thermal backgrounds \cite{shi22prl}, high-energy tails \cite{tomita20jpd,beausang11rsi}, and wave activities \cite{sakai20pop,sakai23pop}. 

Because the signal intensity of Thomson scattering in a single spectrometer channel is proportional to the phase-space density of electrons at the measured wavelength, the intensity is low in high-temperature and low-density plasmas. 
In magnetically confined plasmas, Thomson scattering measurement is one of the most reliable diagnostics to measure the spatiotemporal evolution of electron temperature and density, despite the limited signal-to-noise ratio in high-temperature plasmas \cite{narihara95rsi,narihara01rsi,pasqualotto04rsi,van-der-meiden06rsi,lee10rsi,pasch16rsi}. In conventional Thomson scattering systems in magnetically confined plasmas, such as that in the CHS \cite{narihara95rsi} and the LHD \cite{narihara01rsi} with filtered polychrometers, the number of wavelength channels is limited to $\sim 10$. One of the reasons for the limited number of wavelength channels is the weak Thomson scattering signal compared to the background emissions and noises. 
We obtain the scattered spectrum by making a histogram of each scattered photon whose wavelength depends on the velocity of an electron that interacts with the incident photon. The scattered spectrum has unavoidable statistical noise due to the stochastic nature of scattering. Since the number of detected photons is small in the conventional systems, it is reasonable to reduce the statistical noise by increasing the wavelength integration width to accumulate the scattered photons, in other words, the bin width of the histogram. As a result, it is difficult to distinguish the shape of electron velocity distribution functions because of the limited wavelength channels. 

In addition to the spectral resolution, the temporal resolution is essential because the impulsive energy release events transiently distort distribution functions which subsequently relax into Maxwellian distribution functions on short timescales. One of the relaxation processes is the binary collision between particles. The timescale of electron-electron and electron-ion collisions is approximately $\tau \sim  \SI{20}{\micro s} (T_e/\SI{1000}{eV})^{1.5} (n_e/\SI{e19}{m^{-3}})^{-1}$ with an electron temperature of \SI{1000}{eV} and an electron density of \SI{e19}{m^{-3}}, \cite{spitzer06} therefore, the measurement timescale or integration time should be less than the collision timescale to observe the transient velocity distribution function associated with the impulsive energy release events such as MHD bursts and magnetic reconnections. For example, MHD bursts impulsively excite large-amplitude waves, deforming distribution function within \SI{100}{\micro s} \cite{ida22cphys}. Such rapid events, as well as the subsequent thermalization, require diagnostics with a fast temporal resolution. The pulse laser with a pulse duration of $\sim \SI{10}{ns}$ is appropriate for high temporal resolution in a single shot. Moreover, fast measurements using a pulse laser and a detector with a short integration time reduce the background emissions, resulting in a better signal-to-noise ratio for non-Maxwellian distribution functions. 

There are measurements to observe the shapes of Thomson scattering spectra using \SI{10}{J}-class energetic probe lasers and imaging spectrometers in magnetically confined plasmas \cite{van-der-meiden06rsi,van-lammeren92nf}, however, the signal is sometimes not detected because the intensity of Thomson scattering is lower than the detection threshold in high-temperature plasmas. In order to observe shapes of electron velocity distribution functions even in high-temperature plasmas for the systematic study of electron kinetics, it is necessary to increase the number of scattered photons reaching the detector. Additionally, it is important to know the shape of scattered spectra in various types of electron velocity distribution functions. Although the shapes of scattered spectra have been investigated using Monte Carlo simulations that directly solve particle trajectories in given electromagnetic fields \cite{pastor12nf}, the complete understanding of Thomson scattering spectra in non-Maxwellian electron velocity distribution functions is still under way. 

In this paper, we discuss the conceptual design of a Thomson scattering spectroscopy system capable of measuring non-Maxwellian electron velocity distribution functions in CHD. Note that the proposed Thomson scattering system is not a replacement for the standard Thomson scattering system with filtered polychromators [e.g., Ref.~\onlinecite{narihara01rsi}] because the repetition rate is much lower than that in the conventional systems. Although conventional systems aim to obtain electron temperature and density accurately with high spatial and temporal resolutions, the proposed system focuses on the shape of the scattered spectrum itself, i.e., non-Maxwellian distribution functions.
Section~\ref{sec:design} describes the design of the Thomson scattering system, especially the setting of the spectrometer with high wavelength resolution. The estimated number of scattered photons is large enough to measure the shape of electron velocity distribution functions. The architecture of the spectrometer is determined so that the electron temperature within 100--\SI{6000}{eV} can be measured. 
In Sec.~\ref{sec:simulation}, we calculate synthetic Thomson scattering spectra to estimate the measurement error using Monte Carlo simulations in terms of photon scattering. The synthetic spectra in non-Maxwellian electron velocity distribution functions are calculated. We show that it is possible to measure non-Maxwellian electron velocity distribution functions using the spectrometer with high wavelength resolution. 
In Sec.~\ref{sec:discussion}, we provide a discussion and summary of our research.

\section{Design and efficiency of Thomson scattering system} \label{sec:design}

\subsection{Scattered spectra with Maxwellian distribution functions} \label{sec:theor_spec}

\begin{figure}
    \centering
    \includegraphics[width=\hsize]{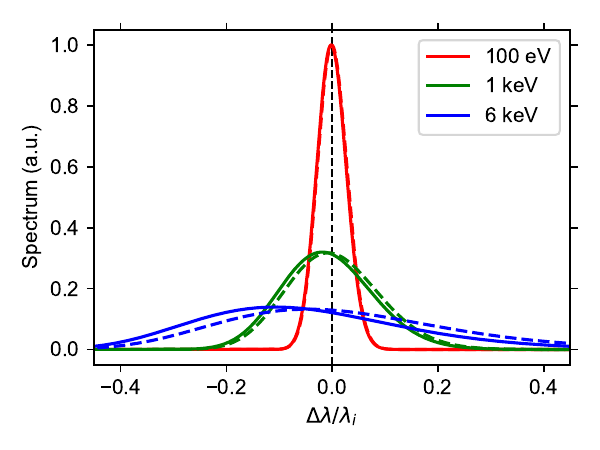}
    \caption{Thomson scattering spectra in Maxwellian distribution functions at $T_e=100$, 1000, and \SI{6000}{eV}.}
    \label{fig:theor_spec}
\end{figure}

First, we consider the shape of scattered spectrum with Maxwellian distribution functions as a reference. Figure~\ref{fig:theor_spec} shows the scattered spectra with different temperatures. We use the same scattering angle of $\theta=\SI{163}{\degree}$ as is adopted in CHS \cite{narihara95rsi}. The horizontal axis is the Doppler-shifted wavelength ($\Delta \lambda$) normalized to the probe laser wavelength ($\lambda_i$). The red, green, and blue curves show the spectra with $T_e=100$, 1000, and 6000~eV, respectively. These temperatures correspond to the expected parameters in CHD, and the deformations from Maxwellian distribution functions are observed within the temperature range in LHD experiments \cite{yamada25jfe}. The solid and dashed curves represent the spectra with and without the relativistic aberration correction, respectively \cite{hutchinson02,williamson71jpp}. Even at the highest temperature, the relativistic blue shift does not appear significantly and the shape of scattered spectra is close to the Gaussian shape. 
The required spectral resolution differs depending on the temperature. To ensure appropriate resolution over the entire temperature range, the wavelength resolution must be determined by the lowest-temperature case, while the wavelength range must be determined by the highest-temperature case. This corresponds to setting a wavelength resolution that divides the Doppler broadening corresponding to \SI{100}{eV} into more than ten bins, while simultaneously covering the Doppler broadening corresponding to \SI{6000}{eV}. The Doppler broadening is $\sim 0.1 \lambda_i$ at \SI{100}{eV}, and $\sim 0.6\text{--}0.7 \lambda_i$ at \SI{6000}{eV}. Therefore, the constraints of the wavelength axis in the spectrometer are that the wavelength resolution is less than $0.1 \lambda_i$ and the entire wavelength range is greater than $0.6\text{--}0.7 \lambda_i$. Therefore, $\gtrsim 100$ channels, which is 10--100 times larger than that in conventional systems \cite{narihara95rsi,narihara01rsi}, are required to measure the shape of electron velocity distribution functions with $100\le T_e\le \SI{6000}{eV}$ using a single spectrometer system. We adopted an imaging spectrometer with gratings to satisfy the above constraints.
Because the required wavelength range increases as a function of the laser wavelength, a shorter-wavelength laser can be easier to design the spectrometer. In this paper, we assume the laser wavelength of \SI{527}{nm}, which corresponds to a frequency-doubled Nd:glass laser.

\subsection{Number of scattered photons} \label{sec:number}

Because the signal level decreases by two orders of magnitude when we increase the channel number by two orders of magnitude, it is essential to improve the signal level and consequently the signal-to-noise ratio in some ways other than increasing the wavelength integration width. A simple solution is to increase the probe laser energy by two orders of magnitude to obtain a higher intensity of scattered light. In this section, we calculate the number of scattered photons on the detector to estimate whether the laser energy is large enough to measure the shape of electron velocity distribution functions. 

\begin{figure}
    \includegraphics[clip,width=\hsize]{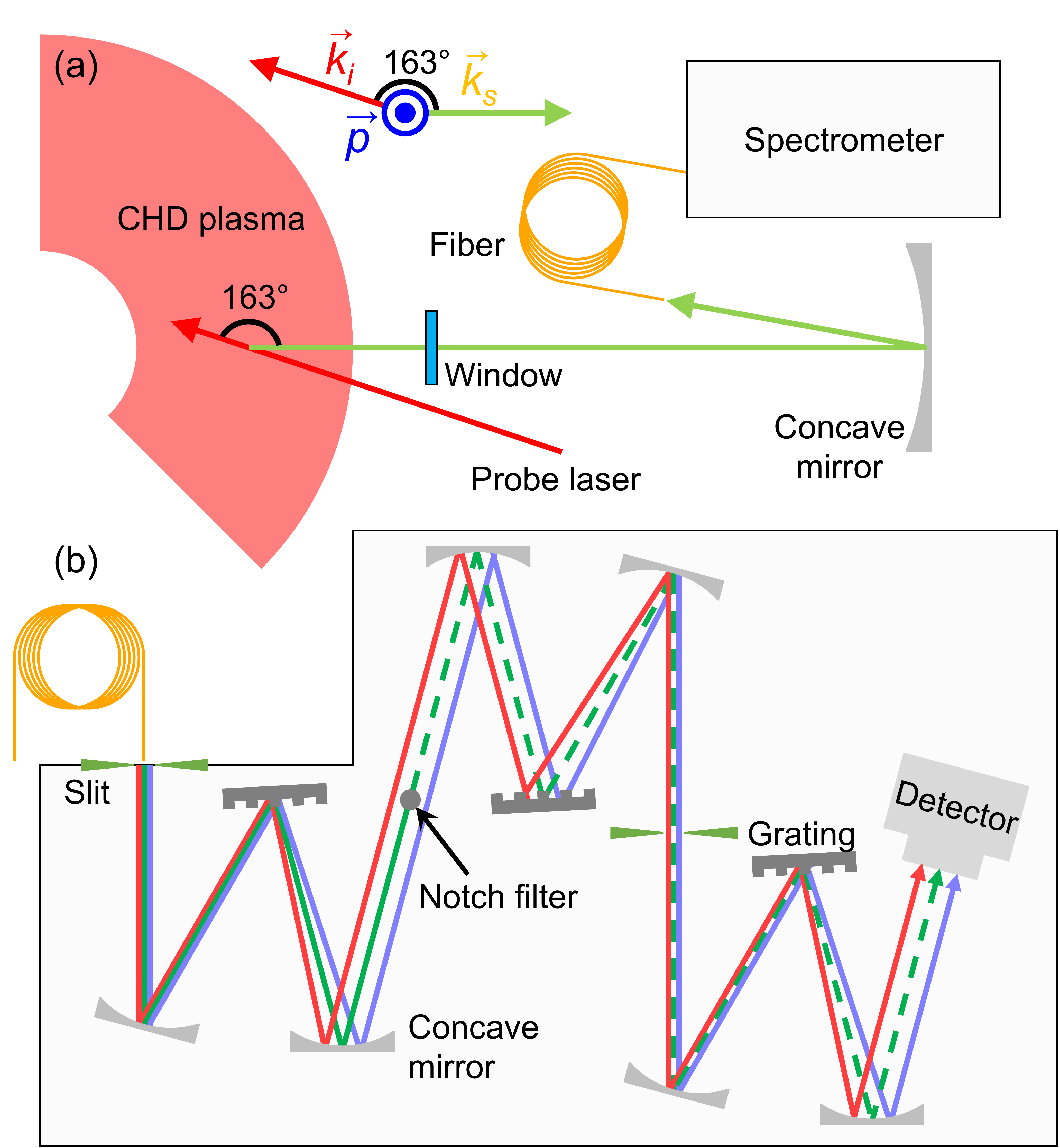}
    \caption{Schematic illustration of (a) the Thomson scattering system and (b) the spatially-resolved spectrometer.}
    \label{fig:design}
\end{figure}

Figure~\ref{fig:design}(a) shows the schematic illustration of the Thomson scattering system in CHD. The torus-shaped CHD plasma is irradiated with a probe laser beam, and the beam is scattered with a scattering angle of $\theta=\SI{163}{\degree}$. We define the wavevector of incident and scattered light as $\vec{k}_i$ and $\vec{k}_s$, respectively. The polarization direction of the probe laser is perpendicular to the scattering plane. The unit vector of the polarization direction $\vec{p}$ is shown in Fig.~\ref{fig:design}(a). The scattered light is collected with a concave mirror with a solid angle of \SI{0.13}{sr}. The reflected light is transferred using a fiber array and goes to a spectrometer. The system is the same as the previous one in CHS \cite{narihara95rsi}, but we replace the probe laser and spectrometer. We consider a triple grating spectrometer with a two-dimensional (2D) image sensor as shown in Fig.~\ref{fig:design}(b), which is similar to that in Ref.~\onlinecite{hassaballa04iiie}. The number of wavelength channels depends on the pixel number of the detector. We assume a camera with an image intensifier as a detector, e.g., Andor iStar sCMOS, Oxford Instruments. Here, we assume that the horizontal pixel number or the number of wavelength channels is 2560. 

We consider the probe laser energy of \SI{75}{J} at \SI{527}{nm}, which is one of the strongest pulse energy commercially available (Elite HE, Amplitude Inc.) The frequency-doubled Nd:glass laser is used as a probe, and the scattered photons are detected with the image intensifier, which has high efficiency at visible wavelengths. The pulse duration is \SI{15}{ns} and the repetition rate is \SI{0.017}{Hz}. Because of the low repetition rate, the Thomson scattering system cannot be a replacement for the standard Thomson scattering system (e.g., Ref.~\onlinecite{narihara01rsi}). We are going to install two lasers to make two measurements in a single discharge. Assuming that the total reflectance and transmittance of upstream optics, which is determined by the mirrors, windows, and polarizers between the laser head and the vacuum vessel, is $\varepsilon_i = 95\%$, the number of photons incident to the plasma is given by the ratio of the total laser energy to the energy of a single photon;
\begin{equation}
    N_i = \frac{E_i \lambda_i}{hc} \varepsilon_i \sim 2\times 10^{20} \frac{E_i}{\text{75 J}} \frac{\lambda_i}{\text{527 nm}} \frac{\varepsilon_i}{95\%},
    \label{eq:photon_i}
\end{equation}
where $E_i$ is the probe laser energy, $h$ is the Planck constant, and $c$ is the speed of light. The incident photons are injected into a plasma and Thomson scattering occurs at a rate determined by the Thomson differential cross section, plasma parameters, and measurement settings. The expected number of scattered photons is given by
\begin{equation}
    N_s = N_i n_e L \frac{d\sigma_T}{d\Omega} \Delta \Omega 
    \sim 2\times 10^{7} \frac{N_i}{2\times 10^{20}}\frac{n_e}{\SI{e19}{m^{-3}}}\frac{L}{\text{1 cm}} \frac{\Delta \Omega}{\text{0.13 sr}} ,
    \label{eq:photon_s}
\end{equation}
where $L$, $d\sigma_T/(d\Omega)$, and $\Delta\Omega$ are the integral length, Thomson differential cross section, and solid angle of the collection optics, respectively. Here we assume that the electron density is \SI{e19}{m^{-3}}, the integral length is \SI{1}{cm}, the solid angle is \SI{0.13}{sr}. We select that the polarization direction of the probe laser beam [$\vec{p}$ in Fig.~\ref{fig:design}(a)] is normal to the scattering plane so that the differential cross section is maximized; $d\sigma_T/(d\Omega) = r_e^2 = \SI{8.0e-30}{m^2~sr^{-1}}$, where $r_e$ is the classical electron radius. Note that the Thomson differential cross section is not constant, especially for high-energy electrons \cite{williamson71jpp}. Here, we use the non-relativistic limit, where the electron speed is much less than the speed of light ($v \ll c$), because the relativistic effect is not significant for electrons with near-zero velocity ($v\sim 0$) and the majority of electrons are $v\sim 0$ in Maxwellian electron velocity distribution functions. This is evident in Fig.~\ref{fig:theor_spec} even with the highest temperature of \SI{6000}{eV} expected in CHD. The correction for energetic electrons is discussed in Sec.~\ref{sec:simulation}. Because the typical electron density in CHD is $n_e \sim (0.2\text{--}5)\times \SI{e19}{m^{-3}}$, the total number of scattered photons is $N_s \sim (0.4\text{--}10) \times 10^{7}$. 

The scattered photon is transferred to the 2D photon detector through a spectrometer, the photon is converted to the photoelectron at a rate determined by quantum efficiency, and finally, the number of photoelectrons on a pixel is recorded as a count. Considering the design discussed in Sec.~\ref{sec:spectrometer}, the detection efficiency is $\varepsilon_d \sim 3\%$ and the estimated number of photoelectrons at the detector is 
\begin{equation}
    N_{e^-} = N_s \varepsilon_d \sim 6\times 10^5 \frac{N_s}{2\times 10^7} \frac{\varepsilon_d}{3\%}.     
    \label{eq:photoelectron}
\end{equation}
When the $6\times 10^5$ photoelectrons are equally divided into 2560 channels on the spectrometer, each channel has more than 200 photoelectrons that are larger than the background emission (see Sec.~\ref{sec:noise}), therefore, the number of detected photoelectrons seems large enough to discuss the shape of electron velocity distribution functions. Note that the estimation of 200 photoelectrons in a single channel is a rough indicator. The detailed analysis of photoelectron number is performed in Sec.~\ref{sec:simulation}. Because the detection efficiency is a function of wavelength, the sensitivity to detect photons is different for different wavelength channels. This effect is also included in the Monte Carlo simulations in Sec.~\ref{sec:simulation}.

\subsection{Design of spectrometer with high wavelength resolution} \label{sec:spectrometer}

\begin{figure}
    \includegraphics[clip,width=\hsize]{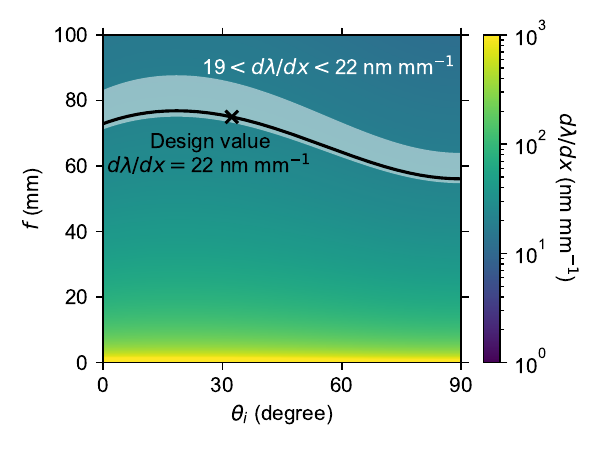}
    \caption{The reciprocal linear dispersion in Eq.~\eqref{eq:RLD} as a function of $\theta_i$ and $f$ with $N_g=\SI{600}{grooves~mm^{-1}}$, $m=1$, and $\lambda=\SI{527}{nm}$.}
    \label{fig:RLD}
\end{figure}

It is essential to reduce stray light for highly accurate and continuous measurements without damaging the detector because the energy of probe laser and consequently the stray light level is much higher than that in the conventional systems. We consider the triple grating spectrometer to reduce the stray light by 8 orders of magnitude \cite{hassaballa04iiie}. As shown in Fig.~\ref{fig:design}(b), this spectrometer consists of three grating spectrometers. The first grating spectrometer adds dispersion and serves as a filter to remove the specific wavelengths by putting notch filter(s) or thin wire(s) at the image plane. These filters physically blocks selected wavelengths to suppress strong signals from stray light and emission lines. The blocked wavelength is determined by the wire position, and the blocking width is proportional to the wire diameter. The notch filter does not affect the wavelength response. The second grating spectrometer adds dispersion opposite to that of the first, thereby converting the light back into white light. The purpose of the second spectrometer is to eliminate stray light originating from surface scattering at the gratings and notch filter(s) in the first spectrometer. After the light is returned to white light, a second slit equivalent to the entrance slit of the first spectrometer is placed at the image plane. Because the surface-scattered light is physically blocked at the second slit, stray light at the laser wavelength is effectively suppressed \cite{hassaballa04iiie}. Finally, the third grating spectrometer adds dispersion again to measure the signal as a spectrum. In order to appropriately cancel the applied dispersion in the second spectrometer, all three spectrometers use the same grating, focusing optics, and incidence angle. 
The wavelength dispersion is determined only by the last grating, last concave mirror, and detector. The grating equation is given by
\begin{equation}
    \sin \theta_i + \sin\theta_d = N_g m \lambda,
    \label{eq:grating}
\end{equation}
where $\theta_i$, $\theta_d$, $N_g$, and $m$ are the incident angle, diffraction angle, groove density, and order of principal maxima, respectively. The wavelength derivative of Eq.~\eqref{eq:grating} gives the diffraction angle change as a function of wavelength. Because the relation between the diffraction angle and the focal position on the detector is given by $dx/d\theta_d = f$, where $f$ is the focal length of the last concave mirror in Fig.~\ref{fig:design}(b), the reciprocal linear dispersion is written as
\begin{equation}
    \frac{d\lambda}{dx} = \frac{d\lambda}{d\theta_d} \frac{d\theta_d}{dx} = \frac{\sqrt{1-(N_g m\lambda - \sin\theta_i)^2}}{N_g m f}. 
    \label{eq:RLD}
\end{equation}
As discussed in Sec.~\ref{sec:theor_spec}, the required wavelength range is $0.6\text{--}0.7 \lambda_i$. Since the detector size is \SI{16.6}{mm}, the required reciprocal linear dispersion is $d\lambda/dx \sim 19\text{--}\SI{22}{nm~mm^{-1}}$.
We plot Eq.~\eqref{eq:RLD} in Fig.~\ref{fig:RLD} with $N_g=\SI{600}{grooves~mm^{-1}}$, $m=1$, and $\lambda=\SI{527}{nm}$. The white region represents the required reciprocal linear dispersion of $19<d\lambda/dx<\SI{22}{nm~mm^{-1}}$. This indicates that the focusing optics with $f=\SI{75}{mm}$ is appropriate to have the designed wavelength range. The reciprocal linear dispersion is not so sensitive to the incident angle. This is technically good in terms of alignment because a small misalignment of the incident angle does not significantly affect the wavelength range. We choose $\theta_i \sim \SI{32}{\degree}$ so that $\theta_i - \theta_d = \SI{45}{\degree}$ for ease of alignment. The reciprocal linear dispersion is $d\lambda/dx \sim \SI{22}{nm~mm^{-1}}$ and the wavelength range is $350 \le \lambda \le \SI{709}{nm}$. The design value and the contour of reciprocal linear dispersion are shown in black marker and curve in Fig.~\ref{fig:RLD}, respectively. 

The current optical design of the Thomson scattering system is summarized below; the collection optics is same as that in CHS \cite{narihara95rsi}. The diameter and curvature radius of the concave mirror in Fig.~\ref{fig:design}(a) are \SI{500}{mm} and \SI{1000}{mm}, respectively. This means the effective focal length is \SI{500}{mm} and F/1. We use a fiber with the core diameter of \SI{200}{\micro m}. The F-number of the fiber is 2. The entrance slit size is assumed to \SI{100}{\micro m}, which is smaller than the fiber core diameter. This difference makes coupling efficiency of fiber 61\%. Within the spectrometer, the signal is transferred by concave mirrors with focal length of \SI{75}{mm}, diameter of \SI{50}{mm}, and F/1.5. We use the same concave mirrors in the spectrometer. While a spectrometer with achromatic lenses is effective and easy to align, it contains finite chromatic aberration, which can be a significant problem in broadband spectroscopy in a single system (e.g., \SI{359}{nm} range in the current design). We choose reflective optics because they do not have chromatic aberration in principle. The use of achromatic and camera lenses will be discussed in a subsequent paper on the detailed design. As mentioned before, three holographic gratings with \SI{600}{grooves~mm^{-1}} and $50\times \SI{50}{mm^2}$ surface area add or subtract the wavelength dispersion in the triple grating spectrometer \cite{hassaballa04iiie}. The F-number of the spectrometer is 1.5 and this is smaller than that of fiber, indicating the loss due to F-number mismatch can be negligible. As discussed in Sec.~\ref{sec:noise}, we put two notch filters or Rayleigh blocks for the laser wavelength and the H$_\alpha$ line. The diameter of the Rayleigh block is comparable to the slit width. The size of intermediate slit is the same as that of the entrance slit. The dispersed signal is obtained by the image sensor with the image intensifier with the pixel size of $6.5\times \SI{6.5}{\micro m^2}$ and $2048\times\SI{2560}{pixels}$. In the current design, the Doppler broadening corresponding to \SI{100}{eV} is divided into $\sim 300$ bins, showing much higher resolution than that in previous measurements \cite{yamada25jfe}. We integrate over $\sim \SI{30}{pixels}$ in space to collect all the signal from a single fiber. The image intensifier operates at the gate mode to reduce background emissions. The gate width of the image intensifier is the same the pulse duration of the probe laser (\SI{15}{ns}). The gain of image intensifier depends on the plasma parameter. At low density and high temperature, the photon number in a channel is less than 100 and the maximum gain of 200 will be applied. On the other hand, at high density and low temperature, the photon number in a channel is $\sim 10000$. Because the pixel well depth of the image sensor is 30000, a gain of several times will be applied. The integration time of the image sensor is \SI{10}{ms}. The instrumental function of the spectrometer is determined by the entrance slit size and is \SI{2.2}{nm}, which is equivalent to $\sim \SI{15}{pixels}$. 

While such imaging spectrometers have both spatial and spectral resolutions, the spatial resolution can be limited because of the light loss in the current setup. Since the purpose of the Thomson scattering system is to measure the shape of electron velocity distribution functions, we plan to start with a single spatial channel and gradually upgrade it to a multi-spatial-channel system.

For the calibration of the system, the relation between wavelength and pixel will be measured using a known light source, such as a neon lamp. The relative wavelength sensitivity will be calibrated by a light source with known continuous spectrum, such as tungsten-halogen lamp. The absolute sensitivity will be calibrated by Rayleigh scattering or rotational Raman scattering \cite{yamada07pfr}. The correction of aspherical aberration will be corrected using a pair of cylindrical focusing optics \cite{kado03jpfr}.

The damage in optics should be avoided during the transport of energetic laser beam, which can destroy the upstream optics if the fluence is higher than the damage threshold. In order to limit the fluence below the damage threshold, the beam diameter should be 50--\SI{75}{mm}. We are currently considering a design where the beam is focused by the focusing optics outside the vessel, resulting in a beam diameter of $\sim \SI{1}{cm}$ inside the vessel. The thermal effect can be neglegible because the repetition rate is \SI{0.017}{Hz}.

\subsection{Efficiency as a function of wavelength} \label{sec:efficiency}

\begin{figure}
    \includegraphics[clip,width=\hsize]{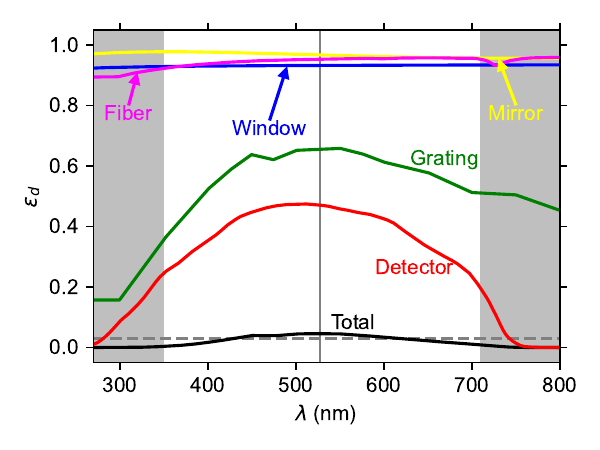}
    \caption{The detection efficiency as a function of wavelength assuming the setup in Fig.~\ref{fig:design}(b).}
    \label{fig:efficiency}
\end{figure}

The efficiency depends on the wavelength, not constant with the wavelength. We estimate the total efficiency at $350\le \lambda\le\SI{709}{nm}$ in Fig.~\ref{fig:efficiency}. The areas not shaded in gray correspond to the wavelength range of the spectrometer. The total efficiency indicated by the black curve includes the efficiency of a detector, three gratings, seven concave mirrors, a window, and a fiber, as shown in Fig.~\ref{fig:design}. In addition to the wavelength response of optical components, we consider the coupling efficiency of fiber (61\%) considered in Sec.~\ref{sec:spectrometer}. The total efficiency is mainly determined by the detector and the gratings. At both ends of the wavelength range, the efficiency drops because the image intensifier in the detector has no sensitivity to both the ultraviolet and infrared wavelengths at \SI{270}{nm} and \SI{750}{nm}. The gray-dashed line represents the 3\% efficiency that we assumed in Sec.~\ref{sec:number}. The efficiency is higher than 3\% around the probe laser wavelength indicated by the vertical line. The detected number of photoelectrons can be larger than the estimate in Eq.~\eqref{eq:photoelectron} if the electron temperature is low because the efficiency in the corresponding wavelength range is higher than the 3\% assumed in Eq.~\eqref{eq:photoelectron}.

\subsection{Background emissions} \label{sec:noise}

In magnetically confined plasmas, there are electromagnetic radiations besides Thomson scattering, e.g., thermal bremsstrahlung and line emission from atoms. These emissions and Thomson scattering signal are observed simultaneously, therefore, these emissions can be one of the sources of photon noise. To distinguish the Thomson scattering spectra from the measured ones, it is necessary to consider the competition of these radiations in the wavelength range of measurement.
The photon spectrum of thermal bremsstrahlung in a unit volume, time, wavelength, and solid angle is given by \cite{rybicki85}
\begin{equation}
    \frac{d N_{B}}{dV dt d\lambda d\Omega} = \frac{32 \pi e^6}{3 m_e  c^3 h\lambda} \sqrt{\frac{2\pi}{3 m_e k_b T_e}} Z_{eff} n_e^2 \exp \left(- \frac{hc}{\lambda k_b T_e}\right) g_{ff},
    \label{eq:brems}
\end{equation}
where $e$, $Z_{eff}\sim 2$, $g_{ff} \sim 1$ are the elementary charge, effective ionization state, and Gaunt factor, respectively. 

\begin{figure}
    \includegraphics[clip,width=\hsize]{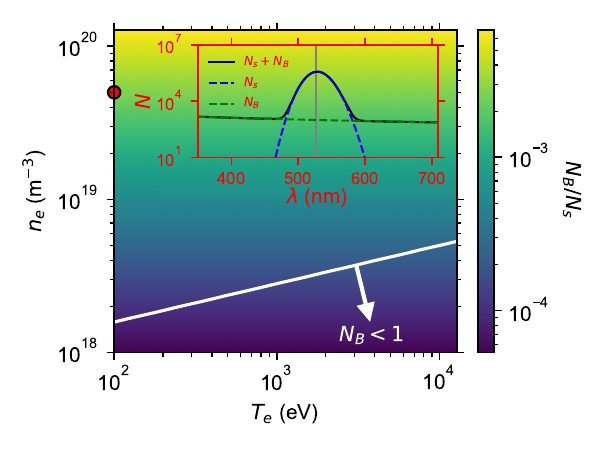}
    \caption{Estimated photon number of thermal bremsstrahlung normalized to Thomson scattering at the laser wavelength as a function of electron temperature and density. The inset shows the photon spectrum of Thomson scattering and thermal bremsstrahlung with $T_e=\SI{100}{eV}$ and $n_e = \SI{5e19}{m^{-3}}$.}
    \label{fig:brems}
\end{figure}

Figure~\ref{fig:brems} shows the ratio of Eq.~\eqref{eq:brems} to Eq.~\eqref{eq:photon_s} in a single wavelength channel of the spectrometer. We assume the volume of $1\times 1\times \SI{40}{cm}$ for thermal bremsstrahlung and $1\times 1\times \SI{1}{cm}$ for Thomson scattering, and the integration time of \SI{15}{ns}, which is the gate width of the image intensifier. 
The difference in the volume originates from the line-of-sight integration length in the imaging system. The integration length in the line-of-sight direction is limited by the probe diameter in Thomson scattering, while the integration length is about the twice of the minor radius in thermal bremsstrahlung. Because we use a concave mirror in an imaging configuration to collect light and couple it into a fiber, signals arriving from positions off the laser beam path (i.e., away from the focal plane) should contribute only weakly, as the coupling efficiency into the fiber decreases. We use an overestimated integration length of \SI{40}{cm} to show that the bremsstrahlung is negligible even with a large measurement volume. 
The ratio is less than $10^{-2}$ and Thomson scattering signal dominates in the parameter space expected in CHD. The white contour shows $N_B=1$. The expected photon number of thermal bremsstrahlung is less than unity with $n_e \lesssim \SI{3e18}{m^{-3}}$ and thermal bremsstrahlung is negligible for low density plasma that is expected to form non-Maxwellian electron velocity distribution function due to less binary collisions. The inset of Fig.~\ref{fig:brems} shows the photon spectrum of Thomson scattering and thermal bremsstrahlung with $T_e=\SI{100}{eV}$ and $n_e = \SI{5e19}{m^{-3}}$, which is indicated by the red marker. The black, blue, and green curves show the total, Thomson scattering, and thermal bremsstrahlung spectra, respectively. The vertical gray line shows the laser wavelength. Although the ratio between thermal bremsstrahlung and Thomson scattering is $\sim 3\times 10^{-3}$ and the thermal bremsstrahlung is negligible, the thermal bremsstrahlung dominates in the tail of the Thomson scattering signal at $\lambda\lesssim \SI{480}{nm}$ and $\lambda \gtrsim \SI{580}{nm}$. If the thermal bremsstrahlung is not properly subtracted from the observed spectrum, this can be identified as an energetic tail of electron velocity distribution function. One should pay particular attention to the analysis of electron velocity distribution function in high density plasmas. 
To accurately subtract signals other than Thomson scattering, the most direct method is to first obtain the emission spectrum without the probe laser, then introduce the probe and measure the difference. By operating the spectrometer with a small time delay, it becomes possible to obtain a spectrum of background radiation alone. Because the image intensifier can acquire two images with a \SI{300}{ns} interval, the background radiation spectrum is obtained \SI{300}{ns} before the laser irradiation. Comparing this data allows for the precise extraction of signals originating solely from Thomson scattering.

The line emission from atoms can be another mechanism that emits light in visible wavelength. The origins of line emissions include impurities, pellets, and neutral beams. For example, high intensity is expected for H$_\alpha$ (\SI{656}{nm}) and H$_\beta$ (\SI{486}{nm}) lines of hydrogen, and C III (\SI{465}{nm}). 
Since the Doppler broadening of H$_\alpha$ line with \SI{1000}{eV} hydrogen is \SI{1.6}{nm} (FWHM), the emission line appears just for $\sim \SI{10}{channels}$, which is much smaller than the total channel number. 
Therefore, when the spectral resolution is high, even considering the Doppler broadening of line emissions, each spectral line is measured over only $\sim 10$ channels. The remaining channels contain no line emissions and instead measure the continuum spectra, such as Thomson scattering and bremsstrahlung emissions.
In the present setup, we assume a spectroscopic system with 2560 channels. Even if about ten spectral lines are observed, $\sim 2400$ channels remain unaffected by line emissions and can be used to measure the scattered light. This means that, when the number of channels is large, contamination from line spectra can be neglected in the data analysis.
The strong hydrogen H$_\alpha$ line coming from the edge region is cut by notch filters as illustrated in Fig.~\ref{fig:design}(b).

\section{Synthetic spectrum}
\label{sec:simulation}

Although the estimated number of photoelectrons on detector in Eq.~\eqref{eq:photoelectron} is much larger than unity, it is still unknown what is the accuracy of estimating electron velocity distribution function. In this section, we performed Monte Carlo simulations to obtain synthetic photon spectra of Thomson scattering in the spectrometer setup discussed in Sec.~\ref{sec:design}. 
The idea of simulating Thomson scattering spectra is to treat the individual scattering process between a single photon and electron and sum up all the photons and electrons. 
Since the expected number of scattered photons in the interaction between a single photon and electron is the case of Eq.~\eqref{eq:photon_s} with $N_i=1$ and $N_e = n_e SL = 1$, where $N_e$ is the total number of electrons within the scattering volume and $S$ is the cross section of probe beam, the total expected number of scattered photons is given by
\begin{equation}
    N_{e^-} = \sum_{j}^{N_i} \sum_{k}^{N_e} \frac{\Delta \Omega }{S} \frac{d^2\sigma_T^{rel}}{d\omega_s d\Omega} \varepsilon_{d}.
    \label{eq:photon_sim}
\end{equation}
Here we replace the differential cross section of Thomson scattering in Eq.~\eqref{eq:photon_s} with that with relativistic correction which is given by \cite{hutchinson02,williamson71jpp}
\begin{equation}
    \frac{d^2\sigma_T^{rel}}{d\omega_s d\Omega} = \frac{r_e^2}{\gamma^2} \frac{1-\vec{\beta}\cdot\vec{i}}{(1-\vec{\beta}\cdot \vec{s})^2} \left|1-\frac{ (1-\cos \theta) (\vec{\beta}\cdot \vec{p})^2}{ (1-\vec{\beta}\cdot \vec{i}) (1-\vec{\beta}\cdot \vec{s}) }\right|^2,
    \label{eq:cross_section}
\end{equation}
where $\gamma = (1-\beta^2)^{-0.5}$ is the Lorentz factor. The electron velocity is different for individual electrons (subscript $k$). Because the differential cross section is a function of electron velocity, this term changes for each electron. 

Using the typical beam diameter of \SI{1}{cm}, the total expected number of scattered photons is $\sim 10^{13}$ and the summation of Eq.~\eqref{eq:photon_sim} is too huge to be calculated even on modern computers. We assume the interaction between ``macro''-photon and ``macro''-electron, which consist with $10^{20}$ photons and $10^{7}$ electrons, respectively, to reduce computational resources. The similar concept is used in particle-in-cell simulations. Because the summand in Eq.~\eqref{eq:photon_sim} is $\sim 6\text{--}7\times 10^{-28}$, the expected number of scattered photons is 0.6--0.7, which is still less than unity and does not exceed unity even with a high speed electron. We assume that the multiple scattering is negligible because the probability of Thomson scattering for a single photon and a single electron is much less than unity. 

In order to obtain a synthetic scattered photon spectrum, we generate three random numbers that correspond to three velocity components of a single macro-electron and calculate the estimated number of photoelectron using Eqs.~\eqref{eq:photon_sim} and \eqref{eq:cross_section}. The velocity is sampled from the electron distribution functions: a single Maxwellian, bi-Maxwellian, and kappa distribution functions. The estimated number is compared with a uniform random number between 0 and 1. If the uniform random number is less than the estimated number, the electronic velocity is recorded.
Repeating this process for all the macro-electrons, we have the information of the electron velocity possessed by all the detected photoelectrons. The velocity is converted to the scattered wavelength using the Doppler shift condition 
\begin{equation}
    \frac{\Delta \lambda}{\lambda_i} = \frac{\vec{\beta} \cdot (\vec{i}-\vec{s})}{1-\vec{\beta}\cdot\vec{i}},
    \label{eq:doppler}
\end{equation}
where $\vec{\beta} = \vec{v}/c$ is the electron velocity normalized to the speed of light, $\vec{i} = \vec{k}_i/k_i$ and $\vec{s} = \vec{k}_s/k_s$ shows the unit vector of the propagation direction of incident and scattered light, respectively. 
The wavelength spectrum of scattered photons is generated by making a histogram of photoelectrons in terms of wavelength and dividing by the detection efficiency in Fig.~\ref{fig:efficiency}. We confirmed that the result remains unchanged with different weights of macro-electrons. The statistical noise is estimated by the square root of photoelectron number in a single wavelength bin of the histogram.

\subsection{Signal-to-noise ratio} \label{sec:sn}

\begin{figure*}
    \includegraphics[clip,width=\hsize]{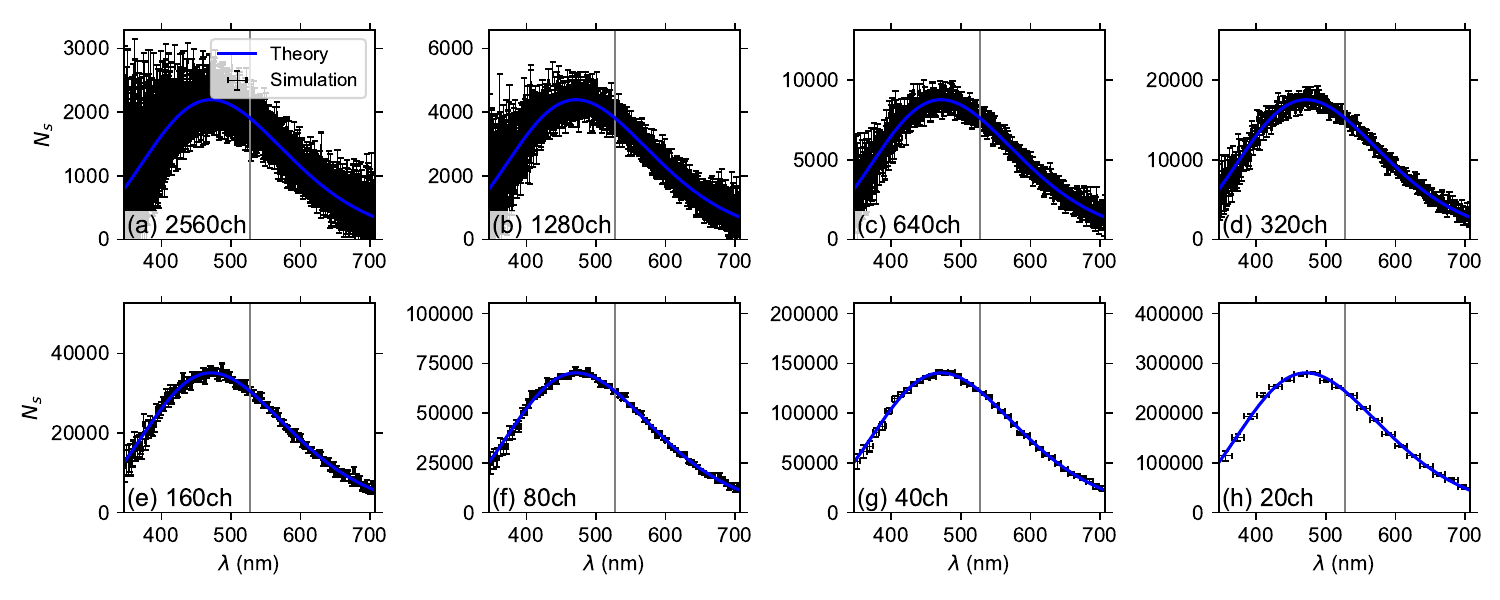}
    \caption{An example of synthetic scattered photon spectrum in a Maxwellian electron velocity distribution function with $T_e=\SI{6000}{eV}$ and $n_e=\SI{2e18}{m^{-3}}$. (a) full resolution spectrum with 2560 channels. (b)--(h) downsampled spectrum of (a).}
    \label{fig:maxwell}
\end{figure*}

\begin{table}
\centering
\caption{Noise sources and estimated counts in a single channel.}
\begin{tabular}{lcc} \hline \hline
Noise source &Characteristics         &Estimated count            \\\hline
Poisson      &$\sigma_p = \sqrt{N_{e^-}}$ &Depends on $T_e$ and $n_e$ \\
Readout      &$\sigma_r = 2.5$        &2.5                        \\
Dark current &$\sigma_d = 0.18 t$     &\SI{1.8e-3}{} \footnote{With \SI{10}{ms} integration time of the image sensor.} \\
EBI          &$\sigma_d = 0.1 t$      &\SI{1.5e-9}{} \footnote{With \SI{15}{ns} integration time of the image intensifier.}       \\\hline \hline
\end{tabular}
\label{tab:noise}
\end{table}

In the detector system considered here, there are some sources of noises: read-out noise ($\sigma_r$), dark current noise ($\sigma_d$), EBI noise ($\sigma_e$), and Poisson noise of photons ($\sigma_p = \sqrt{N_{e^-}}$). We summarize the noise amount in Table~\ref{tab:noise}. The total noise amount in units of photoelectrons in a single channel can be written as $\sigma_{e^-} = (N_s + \sigma_{r}^2+\sigma_{d}^2+\sigma_{e}^2)^{0.5}$. In our case, the most significant noise source can be the Poisson noise of photons, as the observed photon number is small in a single channel. In the Monte Carlo simulations, we can take the Poisson noise into account, which depends on parameters as shown in Table~\ref{tab:noise}. The total noise can be written as a root mean square of each noises. In this section, we investigate the signal-to-noise ratio of scattered spectra with Maxwellian electron velocity distribution functions. 
We generate a Gaussian random numbers which correspond to the electron velocity of Maxwell electron velocity distribution function to calculate synthetic spectra,
\begin{equation}
    f_e(\vec{v})d^3v = (\pi v_{th}^2)^{-1.5} \exp \left( - \frac{v^2}{v_{th}^2} \right) d^3v.
    \label{eq:maxwell}
\end{equation}

Figure~\ref{fig:maxwell} shows an example of synthetic scattered spectra in a Maxwellian electron velocity distribution function with $T_e=\SI{6000}{eV}$ and $n_e=\SI{2e18}{m^{-3}}$, which is expected to be the noisiest in the expected parameters in CHD. The horizontal and vertical axes are wavelength and photon number, respectively. The direct outputs from the simulation are the detected photoelectron spectra, $N_{e^-} (\lambda)$, and the noise expressed in units of photoelectrons, $\sigma_{e^-}(\lambda)$. To make the spectral shape easier to interpret, we divide $N_{e^-} (\lambda)$ and $\sigma_{e^-}(\lambda)$ by the wavelength response function shown in Fig.~\ref{fig:efficiency}, and convert the vertical axis from photoelectron spectra to photon spectra. The black markers and blue curves show the simulated and theoretical spectra, respectively. The peak of scattered spectra is located at $\lambda <\lambda_i$ because of relativistic effect. The laser wavelength is indicated by vertical gray line. The horizontal error bar corresponds to the wavelength integration width. 
Figure~\ref{fig:maxwell}(a) is the full resolution spectrum with 2560 channels and Figs.~\ref{fig:maxwell}(b)--\ref{fig:maxwell}(h) show downsampled spectra using pixel binning. After binning $N$ channels, the total noise becomes $\sigma_{e^-} = [\sum_n^N N_{e^-,n} + N(\sigma_{r}^2+\sigma_{d}^2+\sigma_{e}^2)]^{0.5}$, where $N_{e^-,n}$ is the number of photoelectrons in the $n$-th channel. The error in vertical axis is high in the long and short wavelength regions due to the low efficiency in Fig.~\ref{fig:efficiency}. As the number of channels decreases, the error in vertical axis normalized to the signal intensity decreases because $N_{e^-}$ increases and the Poisson noise of photoelectrons is defined as $\sigma_p = \sqrt{N_{e^-}}$. The simulated spectra with small channel number are in good agreement with the theoretical spectra. This suggests that the post-process pixel binning can dramatically reduce noise even if noise is high at the highest resolution.
Because an appropriate wavelength resolution should exist depending on the shape of distribution functions, the optimal number of channels is determined by the trade-off between signal-to-noise ratio and wavelength resolution.
In this design, we aim to construct a system capable of measuring various non-Maxwellian distribution functions. Therefore, data will be acquired at the highest wavelength resolution, and pixel binning will be applied during data analysis to achieve an appropriate signal-to-noise ratio. This approach is more flexible than fixed $\sim 10$ channel filtered polychromator system for the measurement of non-Maxwellian distribution functions especially in low temperature because most scattered photons are detected in a few wavelength channels with $\sim 10$ channel system.

\begin{figure*}
    \includegraphics[clip,width=\hsize]{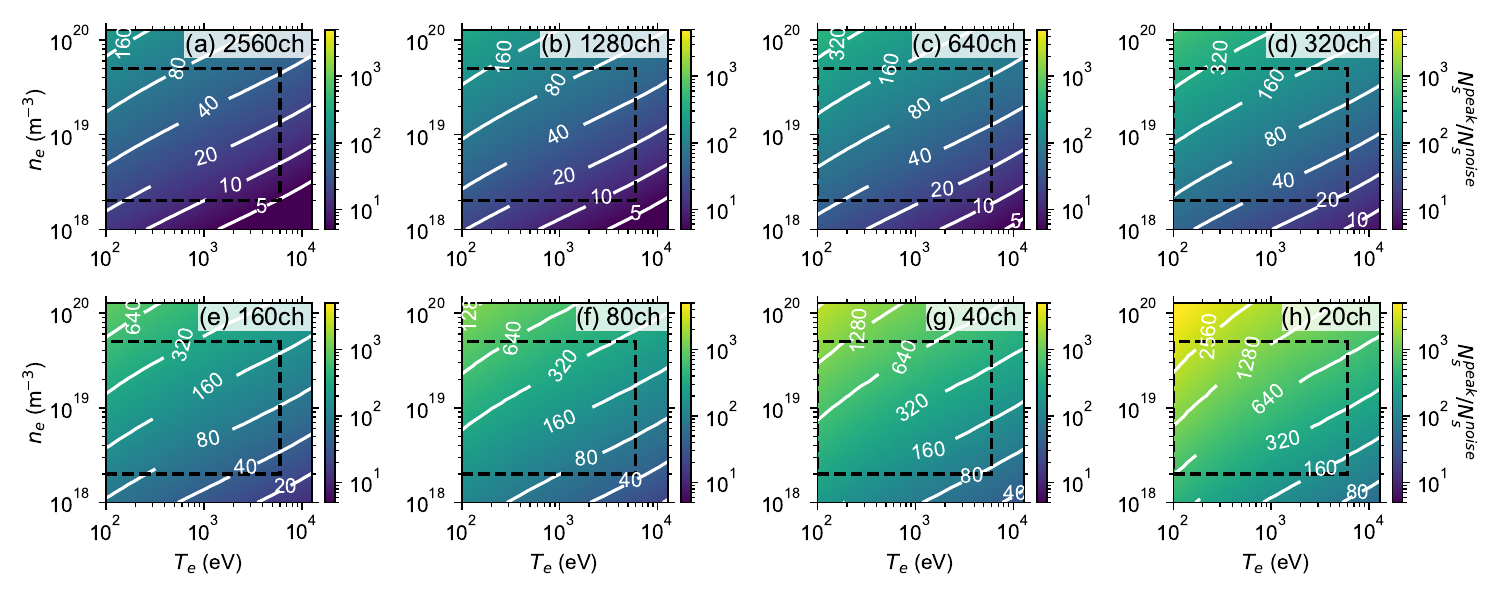}
    \caption{(a)--(h) Signal-to-noise ratio at the peak of scattered spectrum as a function of electron temperature and density for 2560--20 channels, respectively.}
    \label{fig:sn}
\end{figure*}

Figure~\ref{fig:sn} shows the parameter dependence of signal-to-noise ratio. We define the signal-to-noise ratio as the ratio of theoretical photon number to the amplitude of noise at the peak of scattered spectrum. The horizontal and vertical axes are the electron temperature and density, respectively, and the color shows the signal-to-noise ratio. The expected parameter region in CHD is indicated by the black-dashed rectangle. Figures~\ref{fig:sn}(a)--\ref{fig:sn}(h) correspond to 2560--20 channels, respectively. In order to obtain this figure, the synthetic spectra are calculated with various plasma parameters, $100 \le T_e \le \SI{12800}{eV}$ and $1.0\times10^{18}\le n_e \le \SI{1.28e20}{m^{-3}}$, and 5000 independent random seeds for each parameter. The signal-to-noise ratio is low ($\sim 5$) for high temperature and low density, which is better condition to observe non-Maxwellian electron velocity distribution function because of less binary collisions. However, the post-process binning dramatically enhance the signal-to-noise ratio and the ratio becomes more than 100 with 20 channels with all the expected parameters in CHD. Measuring the shape of electron velocity distribution function requires large channel number, therefore, there should be an appropriate channel number and signal-to-noise ratio to measure non-Maxwellian electron velocity distribution function. Note that reducing the number of channels too much may result in poor parameter estimation because the wavelength integration width is comparable to the spectral width of Thomson scattering especially at low temperatures.

\begin{figure}
    \centering
    \includegraphics[clip,width=\hsize]{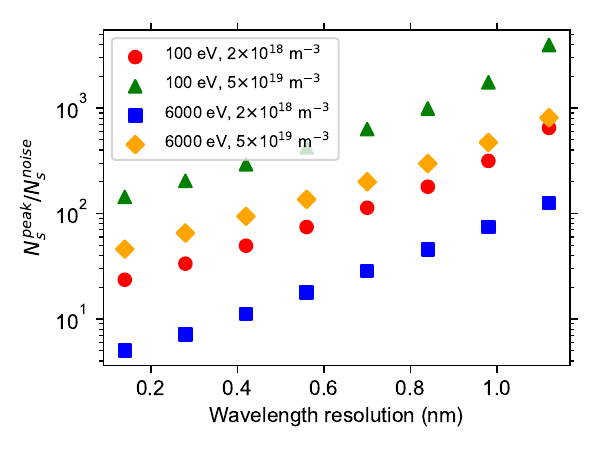}
    \caption{Signal-to-noise ratio as a function of wavelength resolution.}
    \label{fig:binning}
\end{figure}

Figure~\ref{fig:binning} shows the signal-to-noise ratio as a function of channel number or the wavelength resolution. We plot four parameters that are four edges of the black-dashed rectangle in Fig.~\ref{fig:sn}. As the resolution becomes low, the signal-to-noise ratio increases. For the distribution function measurements, both the wavelength resolution and signal-to-noise ratio are the key parameters and both should exceed a threshold, which depends on the shape of distribution functions, to enable the measurement.

\subsection{Errors in estimated parameters} \label{sec:estimate}

\begin{figure}
    \includegraphics[clip,width=\hsize]{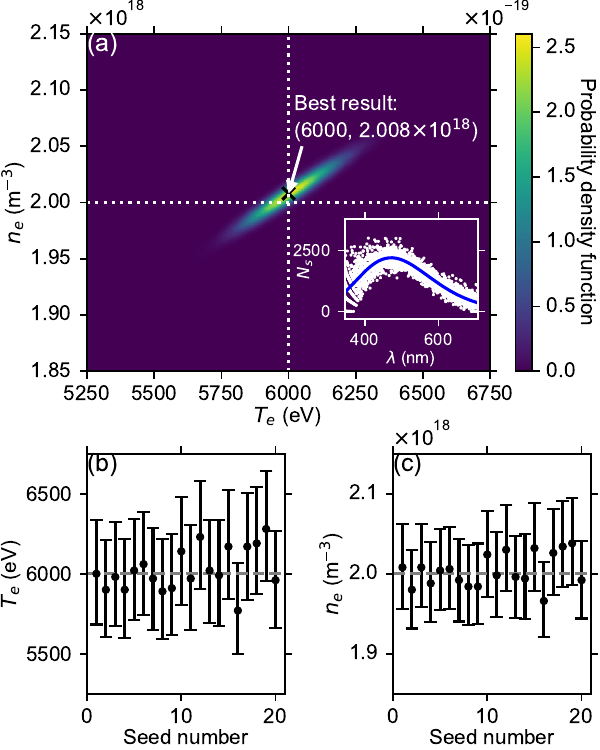}
    \caption{Bayesian inference of electron temperature and density assuming a single Maxwellian distribution function. (a) Posterior probability density function as a function of electron temperature and density. The estimated (b) temperature and (c) density for synthetic spectra with different random seeds.}
    \label{fig:bayes}
\end{figure}

In Sec.~\ref{sec:sn}, we discuss the noise level using the true temperature and density. However, especially if the signal is noisy, the estimated value is sometimes different from the true one. In this section, we discuss the errors in the estimated parameters in Maxwellian electron velocity distribution functions. 
In order to show the noisiest case, we employ the full resolution spectrum with 2560 channels, the highest electron temperature of \SI{6000}{eV}, and the lowest electron density of \SI{2e18}{m^{-3}} expected in CHD. We estimate the electron temperature and density using Bayesian inference. We use the population annealing Monte Carlo sampling method implemented in ODAT-SE (formally known as 2DMAT) \cite{motoyama22cpc}. We minimize the squared residuals between simulated and theoretical photoelectron spectra, not scattered photon spectra, because the noisy signal at both ends of the spectrum becomes large due to the response function in Fig.~\ref{fig:efficiency} and the both ends will be overfitted. The comparison between the spectral shapes in simulation and theory gives the electron temperature. The electron density is estimated globally over all wavelength channels. The conversion factor between the number of photoelectrons and the electron density was obtained by comparing Monte Carlo simulations, performed with an artificially high density, with theoretical spectra. The analysis is performed using 20 synthetic spectra with different random seeds to show the dependence on randomness. 

Figure~\ref{fig:bayes}(a) shows an example of posterior probability density function as a function of electron temperature and density. The true values are indicated by white dotted lines, and the best result of estimated parameters, $(T_e,n_e)=(\SI{6000}{eV}, \SI{2.0e18}{m^{-3}})$, is shown in the black marker. Although the best result is slightly different from the true values, the probability density function at the true values is not so different from that at the best result. The inset of Fig.~\ref{fig:bayes}(a) shows the simulated (white marker) and theoretical (blue curve) spectra at the best result. At both ends of spectrum, a few photoelectrons are observed in a single channel of the detector due to low detection efficiency as shown in Fig.~\ref{fig:efficiency}, therefore, the simulated signal is sometimes much higher than the theoretical one and this results in higher temperature than the true value. When the temperature becomes high, the number of electron within the range of a single wavelength channel decreases as indicated by Eq.~\eqref{eq:maxwell}, resulting in less scattered intensity. In order to compensate the less intensity, the estimated density increases. This relation is shown in the probability density function in Fig.~\ref{fig:bayes}(a); the higher the estimated temperature, the higher the estimated density. 

We compare the estimated temperature and density with the different synthetic spectra in Figs.~\ref{fig:bayes}(b) and \ref{fig:bayes}(c), respectively. The horizontal axis shows the serial number of the random seeds. The first seed corresponds to Fig.~\ref{fig:bayes}(a). The horizontal gray-dashed line shows the true value. The error bar is determined by the 95\% confidence interval or twice of the standard deviation. 
The true values lie within the error bar, and the estimation of electron temperature and density also works with 2560 channels. This indicates that we can observe the shape of scattered spectra even in high-temperature plasmas.

\subsection{Non-Maxwellian electron velocity distribution functions} \label{sec:nonMaxwell}

\begin{figure}
    \includegraphics[clip,width=\hsize]{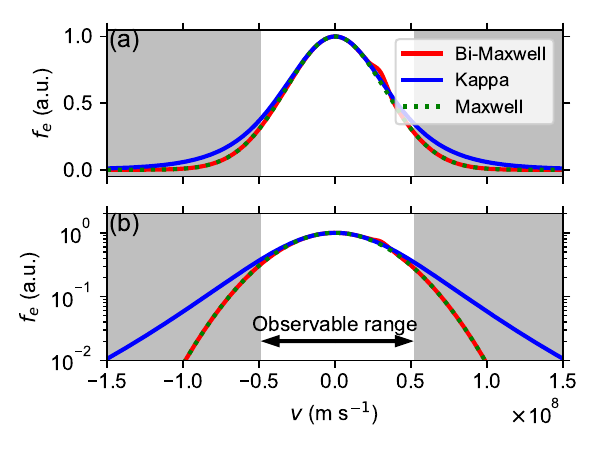}
    \caption{The electron velocity distribution functions on (a) linear and (b) logarithmic scales.}
    \label{fig:edf}
\end{figure}

Due to the large number of channels and large signal-to-noise ratio, it is possible to investigate non-Maxwellian electron velocity distribution function using the Thomson scattering system discussed here. In this section, we simulate scattered spectra with bi-Maxwellian and kappa distribution functions, which are examples of non-Maxwellian distribution functions with wave-particle interaction and energetic electrons respectively \cite{izacard16pop,pierrard10sp}, and confirm whether differences from a single Maxwellian distribution function can be observed. 

For bi-Maxwellian electron distribution function, we assume a high-temperature and high-density thermal component at rest and a moving component with low temperature and density, which is referred to as beam component. We use the  thermal component with \SI{6000}{eV} and \SI{2e18}{m^{-3}}, and the beam component with \SI{100}{eV} and \SI{2e16}{m^{-3}} moving at \SI{3e7}{m~s^{-1}}. This is shown in the red curve in Fig.~\ref{fig:edf}. 
A kappa distribution function is defined by \cite{pierrard10sp}
\begin{equation}
    f_e(\vec{v}) dv^3 = \frac{1}{(\pi \kappa w_{\kappa}^2 )^{1.5}} \frac{\Gamma (\kappa+1)}{\Gamma (\kappa-0.5)} \left( 1+\frac{v^2}{\kappa w_{\kappa}^2} \right)^{-(\kappa+1)} d^3v,
    \label{eq:kappa}
\end{equation}
where $\kappa$ is the spectral index, $w_\kappa = \sqrt{(2\kappa-3)k_b T_e / (\kappa m_e)}$ is the most probable speed, and $\Gamma$ is the gamma function. We use a $\kappa=3$ kappa distribution function with \SI{6000}{eV} thermal component, which is shown in the blue curve in Fig.~\ref{fig:edf}. As a reference, we plot a single Maxwellian electron distribution function as a green dotted curve in Fig.~\ref{fig:edf}. The observable velocity range of the spectrometer is shown in white. The random number of kappa distribution function is generated by the method in Ref.~\onlinecite{zenitani22pop}.
Note that the specific parameters are selected to demonstrate the sensitivity of the spectral shape to non-thermal features and there are no physical justifications on the parameters above. We model the distorted distribution function and the superthermal tail with bi-Maxwellian and the kappa distribution, respectively, and the essence lies in whether characteristic deviations from Maxwellian distribution functions are measurable. Whether such distribution functions are expected to form remains future work, including experiments and simulations.

\begin{figure*}
    \includegraphics[clip,width=\hsize]{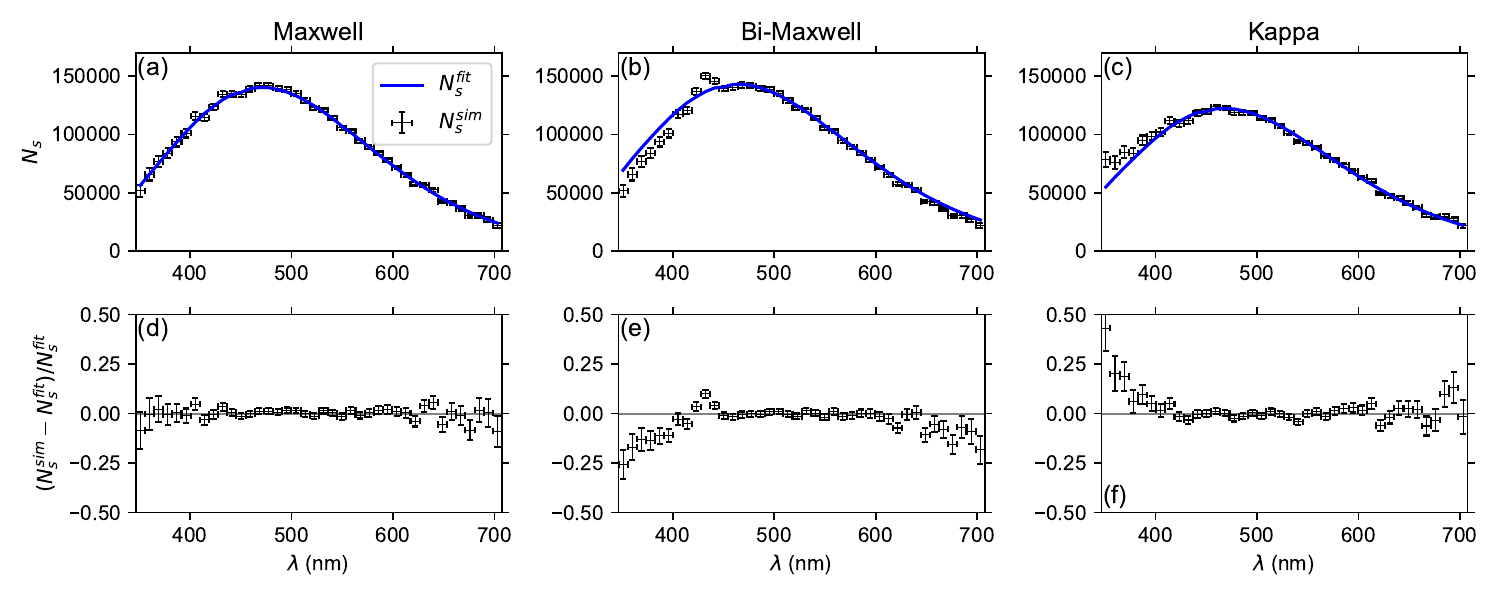}
    \caption{(a)--(c) Simulated spectra of Maxwellian, bi-Maxwellian, and kappa distribution functions in Fig.~\ref{fig:edf} with $n_e=\SI{2e18}{m^{-3}}$, respectively. (d)--(f) Difference of simulated spectra and Maxwellian fits.}
    \label{fig:nonmaxwell}
\end{figure*}

Figure~\ref{fig:nonmaxwell} shows the simulated spectra of Maxwellian, bi-Maxwellian, and kappa distribution functions in Fig.~\ref{fig:edf} with $n_e=\SI{2e18}{m^{-3}}$. The black markers in Figs.~\ref{fig:nonmaxwell}(a)--(c) is the simulation results ($N_s^{sim}$) with 40 channels. This is fitted with the theoretical spectrum with a single Maxwellian electron distribution function that is shown in blue curve ($N_s^{fit}$). The fitting results are $(T_e,n_e)=(\SI{6100}{eV},\SI{2.0e18}{m^{-3}})$ for Maxwellian, $(\SI{6900}{eV},\SI{2.2e18}{m^{-3}})$ for bi-Maxwellian, and $(\SI{6800}{eV},\SI{1.8e18}{m^{-3}})$ for kappa distribution function. Comparing the simulated and fitted spectra, one can find some differences, e.g., a bump at $\lambda\sim \SI{430}{nm}$ in Fig.~\ref{fig:nonmaxwell}(b) and a tail at $\lambda\lesssim \SI{420}{nm}$ in Fig.~\ref{fig:nonmaxwell}(c). This difference suggests that the simulated spectrum cannot be explained by the fitting assuming a single Maxwellian electron distribution function. 

We plot the difference of simulated and fitted spectra normalized to the fitted spectrum in Figs.~\ref{fig:nonmaxwell}(d)--\ref{fig:nonmaxwell}(f). In the Maxwellian case, the signal is close to 0 and uniform in wavelength, therefore, the difference is not evident. 
In the bi-Maxwellian case, there are three deviations from 0: a bump at $\lambda\sim \SI{430}{nm}$ and decrease toward both ends of the observed wavelength range. These structures can be explained as the cold moving beam component enlarges the population at the beam velocity while it does not expand the tail of the electron velocity distribution function. 
In the kappa distribution case, the simulated spectrum is stronger than the fitted one at both ends of the observed wavelength range due to large amount of high-energy electrons. In the presence of high-energy electrons, the estimated temperature becomes large, and the estimated density drops because the number of electrons at low speed decreases.

\section{Discussion and summary} \label{sec:discussion}

Because the Doppler broadening is large in higher temperature plasmas, it is reasonable to use near-infrared wavelength without wavelength conversion. In this paper, we assume a frequency-doubled Nd:glass laser to use an image intensifier with high efficiency at the visible wavelength. Another possibility is to use the fundamental wavelength. Without wavelength conversion, the laser energy is $\sim \SI{100}{J}$ and the laser wavelength is \SI{1053}{nm}. According to Eq.~\eqref{eq:photon_i}, the incident photon number is $\sim 3$ times larger than that with the frequency-doubled laser. This is an advantage even taking into account the fact that there are few detectors that can measure the near-infrared region. If a highly sensitive detector for the near-infrared region is developed in the future, a system using a \SI{1053}{nm} laser would also be a possible solution.

To have a high total laser energy, it is possible to use a bunch of burst mode laser pulses \cite{funaba22srep}. If we use the burst mode laser with the repetition rate of \SI{20}{kHz} and the energy of \SI{1}{J/pulse}, it takes \SI{5}{ms} for 100 pulses (or \SI{100}{J}) in total. Because the collisional relaxation timescale is on the order of microseconds, it is difficult to apply the burst mode laser for impulsive non-Maxwellian distribution functions. 
Moreover, the background emission significantly increases because the gate width should be \SI{5}{ms} to obtain all the scattered photons from 100 pulses and the background emission increases by 5 orders of magnitude. Therefore, we choose a \SI{100}{J}-class laser as a probe of the proposed system.

We cannot expect the origin of stray light unless the entire system, not only the Thomson scattering system but also other heatings and diagnostics, is fixed. The CHD project is still under construction, and it is difficult to quantitatively evaluate the amount of stray light.
Since the stray light is expected to strong with the energetic probe laser beam, we show the conceptual design of Thomson scattering system employing the triple grating spectrometer, which shows good performance even in kJ-class laser-plasma experiments with strong stray light \cite{morita13pop}. However, as shown in Fig.~\ref{fig:efficiency}, the three gratings result in low detection efficiency. If the stray light is not so strong, it is possible to use a single grating spectrometer, which has 2--3 times larger efficiency than the triple grating spectrometer, for accurate measurements of Thomson scattering spectra. 

Because the repetition rate is low, the proposed system cannot be a replacement for the conventional polychromator approach. However, this approach can help understanding, for instance, the discrepancy of electron temperature from electron cyclotron emission and Thomson scattering measurements, where non-Maxwell electron distribution functions can be one of the sources of the discrepancy \cite{beausang11rsi}. In the standard polychromator approach, the number of wavelength channels is limited, and the detailed shape of the scattered spectrum is smeared. The proposed system enables us to characterize the shape of electron distribution functions, which tells us whether the discrepancy can be explained by the distortion of distribution functions. Additionally, the proposed system observes both red-shifted and blue-shifted signals, while the standard system does only red-shifted signals. This is suitable to discuss the skewness of the electron distribution function in the presence of directional beams.

We show two examples of non-Maxwellian electron distribution functions in Fig.~\ref{fig:nonmaxwell}. There are many other non-Maxwellian distribution functions than those discussed here. The problem is how to distinguish the types of distribution functions. Since Thomson scattering spectrum is an almost one-dimensional projection of velocity space, the scattered spectrum is sometimes not directly related to the electron velocity distribution function \cite{tomita20jpd}. It is necessary to understand the typical spectrum of many types of distribution functions. Experimentally, the combination of high wavelength resolution measurement presented here with high temporal resolution measurement \cite{funaba22srep} and anisotropy measurement \cite{morita22pre,shi23rsi} will give us constraints to obtain the complete shape of electron velocity distribution functions \cite{kobayashi23pop}.
Another possibility is to use data science. In Bayesian inference, the plausible model for electron velocity distribution function can be obtained by calculating the model evidence \cite{friel12sn}. Comparing the model evidences model by model, we can obtain the most plausible electron velocity distribution function without any preconceptions \cite{motoyama22cpc}.

In summary, we show the conceptual design of Thomson scattering system to measure electron velocity distribution functions in magnetically confined plasmas. The energetic probe laser beam and the high-sensitivity multichannel spectrometer enables us to measure the shape of electron velocity distribution functions with proper signal-to-noise ratio, which is confirmed with the Monte Carlo simulations on Thomson scattering measurement. The simulations suggest that the post-process pixel binning reduces the statistical noise and increases signal-to-noise ratio. Assuming Maxwellian distribution functions, the electron temperature and density can be estimated with high accuracy even in high-temperature plasmas. Non-Maxwellian distribution functions will be identified by comparing observed spectra with their Maxwellian fits.

\begin{acknowledgments}

We appreciate MCPoP project for helpful discussions, S.~Kado for practical comments on the optical design, and H.~Yamaguchi for the discussions on the shape of electron distribution functions. 
This work is supported by JSPS KAKENHI (Grant Numbers 24K17029 
and 23K25859) 
and by the NINS program of Promoting Research by Networking among Institutions (Grant Numbers 01422301 and 01412302).
K.S., T.H., and A.N. are supported by JST (Moonshot R\&D Program) Japan Grant Number JPMJMS24A3 for the data analysis using ODAT-SE. 

\end{acknowledgments}

\bibliography{ref.bib}

\end{document}